\documentclass{acm_proc_article-sp}

\begin{document}
\title{Stochastic Geometry Modeling of Cellular Networks: Analysis, Simulation and Experimental Validation \vspace{-0.35cm}}
\numberofauthors{2}
\author{
% 1st. author
\alignauthor
Wei Lu\\
       \affaddr{Paris-Saclay University}\\
	\affaddr{Laboratory of Signals and Systems (UMR-8506)}\\
       \affaddr{CNRS-CentraleSupelec-University Paris-Sud XI}\\
       \affaddr{3, rue Joliot-Curie}\\
	\affaddr{91192 Gif-sur-Yvette (Paris), France}\\
       \email{wei.lu@l2s.centralesupelec.fr}
% 2nd. author
\alignauthor
Marco Di Renzo\\
       \affaddr{Paris-Saclay University}\\
	\affaddr{Laboratory of Signals and Systems (UMR-8506)}\\
       \affaddr{CNRS-CentraleSupelec-University Paris-Sud XI}\\
       \affaddr{3, rue Joliot-Curie}\\
	\affaddr{91192 Gif-sur-Yvette (Paris), France}\\
       \email{marco.direnzo@l2s.centralesupelec.fr}
}
\maketitle
\begin{abstract}
Due to the increasing heterogeneity and deployment density of emerging cellular networks, new flexible and scalable approaches for their modeling, simulation, analysis and optimization are needed. Recently, a new approach has been proposed: it is based on the theory of point processes and it leverages tools from stochastic geometry for tractable system-level modeling, performance evaluation and optimization. In this paper, we investigate the accuracy of this emerging abstraction for modeling cellular networks, by explicitly taking realistic base station locations, building footprints, spatial blockages and antenna radiation patterns into account. More specifically, the base station locations and the building footprints are taken from two publicly available databases from the United Kingdom. Our study confirms that the abstraction model based on stochastic geometry is capable of accurately modeling the communication performance of cellular networks in dense urban environments.
\end{abstract}
\category{C.2.1}{Network Architecture and Design}{Wireless communication}
\category{C.4}{Performance of Systems}{Modeling techniques}
\category{G.3}{Probability and Statistics}{Stochastic processes}
\category{I.6.4}{Simulation and Modeling}{Model Validation and Analysis}
\category{I.6.5}{Simulation and Modeling}{Model Development.}
\terms{Networks, Modeling, Simulation, Performance Evaluation.}
\keywords{Cellular networks, stochastic geometry, point processes.}
\section{Introduction} \label{Introduction}
Heterogeneous ultra-dense cellular networks constitute an enabling architecture for achieving the disruptive capabilities that the fifth generation (5G) of cellular networks is expected to provide \cite{5G_PPP}. Modeling, simulating, analyzing and optimizing such networks is, however, a non-trivial problem. This is due to the large number of access points that are expected to be deployed and their dissimilar characteristics, which encompass deployment density, transmit power, access technology, etc. Motivated by these considerations, several researchers are investigating different options for modeling, simulating, mathematically analyzing and optimizing these networks. The general consensus is, in fact, that the methods used in the past for modeling cellular networks, e.g., the hexagonal grid-based model \cite{GidBased_Cellular}, are not sufficiently scalable and flexible for taking the ultra-dense and irregular deployments of emerging cellular topologies into account. \vspace{-0.20cm}

Recently, a new approach for overcoming these limitations has been proposed. It is based on the theory of point processes (PP) and leverages tools from stochastic geometry for system-level modeling, performance evaluation and optimization of cellular networks \cite{AndrewsNov2011}. In this paper, it is referred to as the PP-based model. Unlike its grid-based counterpart, the locations of the base stations (BSs) are not assumed to be regularly deployed, but they are assumed to be randomly distributed according to a PP. This approach, due to its mathematical flexibility for modeling heterogeneous ultra-dense cellular deployments, has been extensively used in the last few years and it is gaining exponential prominence in the scientific community. The interested reader is referred to \cite{MDR_TCOMrate}-\cite{MDR_TCOM_EiD} for the latest achievements in this field of research. \vspace{-0.40cm}

Despite that, the experimental validation of the PP-based abstraction for modeling cellular networks has remained elusive to date. This is especially true for modeling macro cellular BSs, whose deployment is, usually, not totally random. A few researchers have tried to justify the PP-based model by using empirical data for the locations of the BSs, e.g., \cite{AndrewsNov2011}, \cite{OpenCellID_PPP}, \cite{Haenggi_Fitting}. These studies have confirmed the potential accuracy and the usefulness of the PP-based model. They, however, are based on a small set of data and on simplifying modeling assumptions. Notably, they do not account for the footprints of the buildings and rely on simplified channel models, where line-of-sight (LOS) and non-line-of-sight (NLOS) propagation conditions that originate from the presence of buildings are neglected. This is mostly due to the inherent difficulties in obtaining accurate data related to the locations of the BSs and of the buildings in urban areas \cite{Heath__Blockage}. The importance of modeling LOS and NLOS propagation conditions, however, has recently been emphasized in several papers, e.g., \cite{LopezPerez__LosNlos} and references therein. \vspace{-0.20cm}

In this paper, we investigate the accuracy of the PP-based abstraction for modeling cellular networks with the aid of experimental data. We explicitly take realistic BS locations, building footprints, LOS/NLOS channel conditions and antenna radiation patterns into account. More specifically: i) the locations of the BSs are taken from a large database made available by OFCOM, the independent regulator and competition authority for the United Kingdom (UK) communications industries \cite{OFCOM}; and ii) the footprints of the buildings are taken from a large database made available by Ordnance Survey, the Britain's mapping agency offering the most up-to-date and accurate maps of the UK \cite{OrdnanceSurvey}. Our extensive study highlights that the PP-based model is capable of accurately predicting the performance of cellular networks in dense urban environments. It also shows, however, that their performance highly depends on the channel models and the antenna radiation patterns being used. \vspace{-0.20cm}

In addition, this paper provides another important contribution: we introduce tractable approximations for modeling empirical LOS/NLOS conditions due to blockages and practical antenna radiation patterns, which facilitate the system-level simulation and the mathematical modeling of cellular networks. The relevance of the proposed approximations is twofold: i) simulation, analysis and optimization of ultra-dense cellular networks turn out to be easier, since neither the actual building footprints nor the actual antenna radiation patterns are needed and, more importantly, ii) they offer a transparent interface between telecommunication operators/manufactures and governmental agencies on one side, as well as academic and research organizations on the other side, since accurate data on the locations of the BSs and on the footprints of the buildings, which often constitutes sensible and confidential information, does not need to be explicitly provided. By using the proposed approximations, this data can be provided in a modified format, which allows the former organizations not to unveil sensible and confidential information on actual network and city deployments. In this paper, we prove that the proposed approximations are accurate enough for modeling cellular networks. In a companion paper, we have recently verified that the proposed approximations are suitable for mathematical modeling and for computing tractable utility functions for system-level optimization as well. Even though a detailed discussion of mathematical modeling is outside the scope of this paper, the interested reader may find preliminary results in \cite{MDR_TWC_mmWave}. \vspace{-0.20cm}

The remainder of this paper is organized as follows. In Section \ref{SystemModel}, the system model is introduced. In Section \ref{Approximations}, the proposed approximations for facilitating system-level simulations and mathematical analysis are presented. In Section \ref{sec_simu_real}, the accuracy of the PP-based abstraction model and of the proposed approximations is substantiated with the aid of experimental data. Also, main findings and takeaway messages are discussed. Finally, Section \ref{Conclusion} concludes this paper.
\section{System Model} \label{SystemModel}
\begin{table}[!t]
\centering
\caption{BS statistics from OFCOM - The city of London ($\mathcal{A} = 4 \, {\rm{km}}^2$).}
\begin{tabular}{|c|c|c|c|} \hline
 &O2+Vod.&O2&Vod.\\ \hline
Number of BSs&319&183&136\\ \hline
Number of rooftop BSs&95&62&33\\ \hline
Number of outdoor BSs&224&121&103\\ \hline
Average cell radius (m)&63.1771&83.4122&96.7577\\
\hline\end{tabular}
\label{Table1} \vspace{-0.25cm}
\end{table}
\begin{table}[!t]
\centering
\caption{BS statistics from OFCOM - The city of Manchester ($\mathcal{A} = 1.8 \, {\rm{km}}^2$).}
\begin{tabular}{|c|c|c|c|} \hline
 &O2+Vod.&O2&Vod.\\ \hline
Number of BSs&37&16&22\\ \hline
Number of rooftop BSs&25&12&14\\ \hline
Number of outdoor BSs&12&4&8\\ \hline
Average cell radius (m)&125.925&191.492&163.305\\
\hline\end{tabular}
\label{Table2} \vspace{-0.25cm}
\end{table}
\subsection{Base Stations Modeling}\label{sec_BS_distribution}
In order to test the accuracy of the PP-based model for the locations of the BSs, we use experimental data from two actual deployments of BSs that correspond to the cities of London and Manchester in the UK. As mentioned in Section \ref{Introduction}, this data is obtained from OFCOM \cite{OFCOM}. More specifically, we consider the BSs of two telecommunication operators: O2 and Vodafone. The empirical data is summarized in Tables \ref{Table1} and \ref{Table2} for London and Manchester, respectively. The following terminology and notation are used: i) ``rooftop BSs'' is referred to the BSs that lay inside a geographical region (polygon) where a building is located; ii) ``outdoor BSs'' is referred to the BSs that lay in a geographical region where no buildings are located. Information about the locations of the buildings is provided in Section \ref{sec_Building_distribution}; and iii) $\mathcal{A}$ denotes the area of the geographical region under analysis. From the number of BSs ($\mathcal{N}$) and $\mathcal{A}$, the density of BSs is obtained as $\lambda_{\rm BS}  = \mathcal{N}/\mathcal{A}$. Accordingly, the average cell radius shown in the tables is computed as $R_c = \sqrt{1/(\pi \lambda_{\rm BS})}$ \cite{MDR_TWC_mmWave}. It is worth mentioning that O2 and Vodafone share one BS in the considered region of the city of Manchester. As expected, the data in Tables \ref{Table1} and \ref{Table2} confirms that a denser deployment of BSs is available in London. \vspace{-0.20cm}

To study the impact of network densification and the potential gains of sharing the BSs between telecommunication operators, two scenarios are investigated. In the first scenario, the BSs of O2 and Vodafone operate at different frequencies, thus they do not interfere with each other. This is equivalent to having just one telecommunication operator in the region of interest. Hence, only the BSs of one telecommunication operator are accessible to the typical mobile terminal (MT). In the second scenario, O2 and Vodafone share the BSs and they operate at the same frequency. So, a denser deployment of BSs is available in the region of interest and the BSs of both telecommunication operators are accessible to the typical MT. Furthermore, full frequency reuse for the BSs of the same telecommunication operator and a saturated load traffic model are assumed. This implies that, with the exception of the serving BS, all the accessible BSs at a given frequency act as interferers for the probe MT. \vspace{-0.20cm}

As far as the PP-based model for the locations of the BSs is concerned, we assume that the BSs are distributed according to a Poisson PP (PPP). It is known, in fact, that the PPP results in a pure random deployment and that it is more mathematically tractable than any other PPs available in the literature \cite{Haenggi_Fitting}. In other words, it corresponds to a worst case scenario for testing the accuracy of the PP-based model.
\begin{figure*}[!t]
\centering
\includegraphics [width=\textwidth] {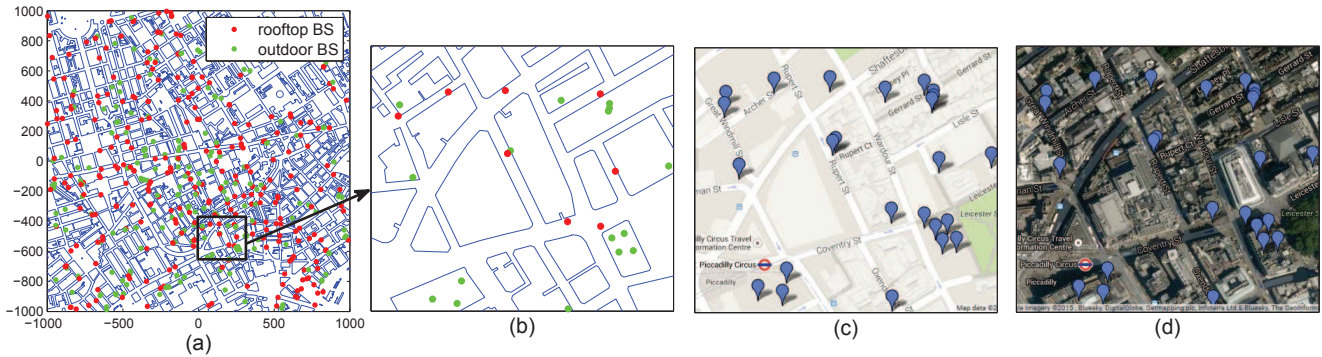}
\vspace{-0.75cm} \caption{London case study. Dense urban environment, where the 55.9\% of the area is occupied by buildings. Horizontal and vertical axis provide distances expressed in meters. (a) Entire region under analysis. (b) Magnification of a smaller region. (c) Google map view of (b). (d) Satellite view of (b).}
\label{Fig_London} \vspace{-0.10cm}
\end{figure*}
\begin{figure*}[!t]
\centering
\includegraphics [width=\textwidth] {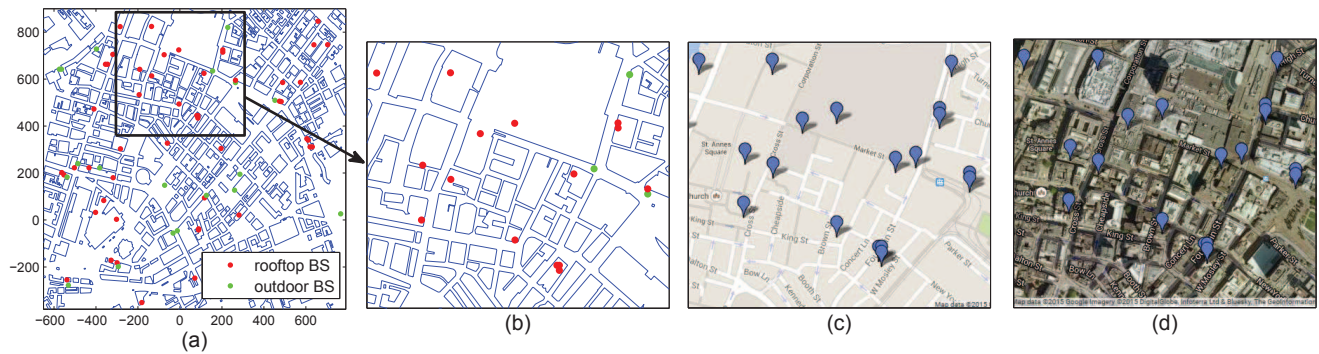}
\vspace{-0.75cm} \caption{Manchester case study. Dense urban environment, where the 44.46\% of the area is occupied by buildings. Horizontal and vertical axis provide distances expressed in meters. (a) Entire region under analysis. (b) Magnification of a smaller region. (c) Google map view of (b). (d) Satellite view of (b).}
\label{Fig_Manchester} \vspace{-0.20cm}
\end{figure*}
\subsection{Buildings Modeling}\label{sec_Building_distribution}
To take realistic blockages into account, i.e., LOS and NLOS propagation conditions due to the locations and the shapes of buildings (see also Section \ref{LOS_NLOS}), we use experimental data corresponding to the actual deployments of buildings in the cities of London and Manchester. As mentioned in Section \ref{Introduction}, this data is obtained from OS \cite{OrdnanceSurvey}. In particular, the same geographical regions as those of the locations of the BSs in Section \ref{sec_BS_distribution} are considered. To make sure that the data obtained from the two independent websites of OFCOM and OS can be merged together, we have verified their consistency with the aid of Google maps for the same areas. Figures \ref{Fig_London} and \ref{Fig_Manchester} provide a graphical representation of the geographical areas under analysis, by merging the data from OFCOM and OS. As far as the buildings are concerned, their elevation is not considered, since this data is not available in the database. Therefore, the analysis of its impact is postponed to a future research study.
\subsection{Blockages Modeling} \label{LOS_NLOS}
The presence of buildings in dense urban environments constitute an inherent source of blockages, which results in LOS and NLOS links. Modeling LOS and NLOS propagation conditions constitute an important requirement for assessing the physical layer performance of transmission schemes within the 3rd generation partnership project (3GPP) \cite{3GPP_pathloss}. As far as the PP-based modeling of cellular networks is concerned, however, this is often overlooked due to the mathematical intractability of modeling links having multiple states. Only recently, in fact, some attempts to take LOS and NLOS links into account have been made, e.g., \cite{MDR_TWC_mmWave} and references therein. In this paper, three approaches for modeling blockages are investigated and compared against each other.
\vspace{-0.20cm}
\subsubsection{Empirical-Based Model} \label{LOS_NLOS__Empirical}
Based on the locations of the BSs and on the locations and the shapes of the buildings described in Sections \ref{sec_BS_distribution} and \ref{sec_Building_distribution}, respectively, LOS and NLOS propagation conditions can be empirically taken into account. In this paper, LOS and NLOS links are identified as follows. Let a generic ``outdoor BS'' and a generic MT in the region of interest. The related link is in LOS if no building is intersected by connecting the BS and the MT with a straight line. Otherwise, the link is in NLOS. Let a generic ``rooftop BS'', the BS-to-MT links are assumed to be in NLOS. This is a simplifying assumption used in other papers as well \cite{Singh__Globe2014}, which seems acceptable if no information on the elevation of the buildings is available.
\vspace{-0.50cm}
\subsubsection{3GPP-Based Model} \label{LOS_NLOS__3GPP}
The 3GPP provides a statistical model for identifying LOS and NLOS links in several scenarios. More specifically, the approach is probabilistic and it does not need any empirical information about the locations of the BSs and of the buildings. On the other hand, a BS-to-MT link is assumed to be in LOS or NLOS with some probability that depends on the BS-to-MT distance. In this paper, we focus our attention only on the MTs that are located in outdoor. This is because differentiating between outdoor and indoor propagation would require to introduce different channel parameters. This generalization is postponed to future research. As a case study, hence, we consider the LOS/NLOS model for outdoor MTs in a urban micro-cell scenario, which provides a good representation of the geographical regions considered for London and Manchester in Tables \ref{Table1}, \ref{Table2} and Figures \ref{Fig_London}, \ref{Fig_Manchester}. In mathematical terms, the probability that a link of length $r$, i.e., the BS-to-MT distance is $r$, is in LOS is \cite{3GPP_pathloss}:
\begin{equation}
\label{3GPP_LOS}
{p^{(\rm LOS)}}\left( r \right) = \min \left\{ {{{18}/ r},1} \right\}\left( {1 - e^{\left( { - {r \over {36}}} \right)}} \right) + e^{\left( { - {r \over {36}}} \right)}
\end{equation}
\noindent where $\min\left\{\cdot,\cdot \right\}$ denotes the minimum function. The probability that the link is in NLOS is ${p^{(\rm NLOS)}}\left( r \right)=1-{p^{(\rm LOS)}}\left( r \right)$.
\vspace{-0.45cm}
\subsubsection{1-State Model} \label{LOS_NLOS__1state}
As mentioned in Section \ref{LOS_NLOS}, LOS and NLOS propagation conditions are often neglected and all links are assumed to be either in LOS or NLOS. This case study is considered in this paper as well, in order to better understand the differences between 1-state and 2-state blockage models. For example, if all the BS-to-MT links are assumed to be in NLOS, ${p^{(\rm NLOS)}}\left( r \right)=1$ and ${p^{(\rm LOS)}}\left( r \right)=0$ regardless of the link length $r$.
\subsection{Channel Modeling}\label{Channel}
In addition to modeling blockages, we consider a practical channel model. In particular, path-loss, shadowing and fast-fading are considered. Let a generic BS denoted by BS$_i$ and a generic outdoor MT denoted by MT$_k$. \vspace{-0.20cm}

As for the path-loss, we consider the physical-based bounded model $l^{\left( S \right)} \left( r_{i,k} \right) = \kappa ^{\left( S \right)} \left( {\max \left\{ {r_0 ,r_{i,k}} \right\}} \right)^{\alpha^{\left( S \right)}}$, where $r_{i,k}$ denotes the BS-to-MT distance, $S={\rm LOS}$ or $S={\rm NLOS}$ if the BS-to-MT link is in LOS or in NLOS, $\kappa ^{\left( S \right)}$ denotes the free-space path-loss, $\alpha ^{\left( S \right)}$ denotes the path-loss exponent, and $r_0$ is a positive constant that avoids the singularity of the path-loss model for $r_{i,k} \to 0$. \vspace{-0.20cm}

As for the shadowing, we consider that it is distributed according to log-normal random variable with mean and standard deviation (in dB) equal to $\mu^{\left( S \right)}$ and to ${\sigma ^{\left( S \right)}}$, respectively. In this paper, shadowing is denoted by $X_{i,k}^{\left( S \right)}$. \vspace{-0.20cm}

As for the fast-fading, we consider that the envelope of the links in LOS and in NLOS is distributed according to a Nakagami-$m$ (with $m>1$) and a Rayleigh random variable with mean power $\Omega$, respectively. In this paper, the envelope of the fast-fading is denoted by $h_{i,k}^{\left( S \right)}$. Fast-fading and shadowing are assumed to be independently distributed. \vspace{-0.20cm}

Thus, the received power at MT$_k$ can be formulated as:
\begin{equation}
\label{Received_power}
{P_R} = {{{P_T}{G_{i,k}}h_{i,k}^{\left( S \right)}{X}_{i,k}^{\left( S \right)}} \over {{\kappa^{\left( S \right)}}{{\left( {\max \left\{ {{r_0},{r_{i,k}}} \right\}} \right)}^{{\alpha^{\left( S \right)}}}}}}
\end{equation}
\noindent where $P_T$ is the transmit power of BS$_i$ and ${G_{i,k}}$ is the antenna gain of the BS$_i$-to-MT$_k$ link. If BS$_i$ is the intended BS of MT$_k$, we assume ${G_{i,k}}=1$. If, on the other hand, BS$_i$ is an interfering BS for MT$_k$, the antenna radiation patterns of BS$_i$ and MT$_k$ are assumed to be randomly oriented with respect to each other and to be uniformly distributed in $\left[ {-\pi,\pi } \right)$. In this case, we have $G_{i,k}  = g_{{\rm{BS}}} \left( {\theta _i } \right)g_{{\rm{MT}}} \left( {\theta _k } \right)$, where $g_{{\rm{BS}}} \left(  \cdot  \right)$, $g_{{\rm{MT}}} \left(  \cdot  \right)$ are the antenna radiation patterns of BSs and MTs, respectively, and ${\theta _i }$, ${\theta _k }$ are the angle off the boresight directions of BSs and MTs, respectively. In this paper, we assume $g\left( \theta  \right)= g_{{\rm{BS}}} \left( \theta  \right) = g_{{\rm{MT}}} \left( \theta  \right) \le 1$ for every $\theta$. Further details about $g\left(  \cdot  \right)$ are provied in Section \ref{Sec_antenna}.
\subsection{Antenna Radiation Pattern} \label{Sec_antenna}
In cellular networks, directional antennas are typically used for enhancing the received power of the intended link and, simultaneously, for reducing the other-cell interference. In this paper, we are interested in understanding their impact on the accuracy of the PP-based model for cellular networks. Two widespread antenna radiation patterns are considered.
\vspace{-0.55cm}
\subsubsection{Omni-Directional Model} \label{Sec_OmniAntenna}
The omni-directional model constitutes the baseline for comparing any other antenna radiation patterns. In particular, $g\left( \cdot  \right)$ can be formulated as follows:
\begin{equation}
\label{G_Omni}
g\left( \theta  \right) = 1 \quad \forall \theta  \in \left[ { - \pi ,\pi } \right)
\end{equation}
\vspace{-0.55cm}
\subsubsection{3GPP-Based Model} \label{Sec_3GPPAntenna}
The 3GPP provides a reference antenna radiation pattern for system-level simulations, which is formulated as \cite{3GPP_pattern}:
\begin{equation}
\label{G_3GPP}
g\left(  \theta  \right) = { \begin{cases}
{{{10}^{ - \frac{3}{{10}}{{\left( {\frac{{2\theta }}{{\theta^{\left( {\rm{3dB}}  \right)}} }} \right)}^2}}}} & \quad \textrm {if}\quad {\left| \theta  \right| \le \varphi }\\
{{{10}^{ - \frac{{{g^{\left(  \min \right) }}}}{{10}}}}} & \quad \textrm {if} \quad {\varphi  < \left| \theta  \right| \le \pi } \\
 \end{cases} }
\end{equation}
\noindent where ${\theta_q^{\left( {\rm{3dB}}  \right)}}$ denotes the 3 dB beamwidth, ${g^{\left(  \min \right) }}$ denotes the minimum gain, and $\varphi  = \theta^{\left( {{\rm{3dB}}} \right)}\sqrt {g^{\left( {\min } \right)}/12}$ denotes the angle that corresponds to the main lobe \cite{3GPP_pattern}.
\subsection{Cell Association Modeling}\label{Sec_association}
The typical (probe) MT is assumed to be served by any accessible BS that provides the highest average received power to it. Thus, path-loss and shadowing are both taken into account for cell association. Fast-fading, on the other hand, is averaged out and neglected. This is, in fact, the typical operating condition based on 3GPP specifications \cite{3GPP_pathloss}. \vspace{-0.20cm}

Let MT$_k$ be the typical MT and the probe link be identified by the subscript ``0''. Let $C_{0,k}^{\left( {S} \right)}$ be defined as follows:
\begin{equation}
\label{Eq_C}
C_{0,k}^{\left( S \right)} = \mathop {\min }\limits_{i \in {\Phi ^{\left( S \right)}_{{\rm{BS}}}}} \left\{ C_{i,k}^{\left(S \right)}={\frac{{{\kappa^{\left( S \right)}}{{\left( {\max \left\{ {{r_0},{r_{i,k}}} \right\}} \right)}^{{\alpha^{\left( S \right)}}}}}}{{{X}_{i,k}^{\left( S \right)}}}} \right\}
\end{equation}
\noindent where $\max\left\{\cdot,\cdot \right\}$ is the maximum function, ${\Phi ^{\left( S \right)}_{{\rm{BS}}}}$ is the PP of the BSs in state $S$, and $1/C_{0,k}^{\left( S \right)}$ is the highest average received power at MT$_k$ from any accessible BS whose BS-to-MT$_k$ link is in LOS if $S=\rm{LOS}$ or in NLOS if $S=\rm{NLOS}$. \vspace{-0.50cm}

From \eqref{Eq_C}, the serving BS of MT$_k$ is that corresponding to the inverse average received power defined as $C_{0,k} =\min \left\{ {C_{0,k}^{\left( \rm LOS \right)},C_{0,k}^{\left( {\rm NLOS} \right)}} \right\}$, since it provides the best link.
\subsection{Problem Statement}
Let BS$_0$ be the serving BS of MT$_k$. Based on \eqref{Received_power} and \eqref{Eq_C}, the signal-to-interference-plus-noise-ratio (SINR) at the typical MT, MT$_k$, can be formulated as follows:
\begin{equation}
\label{Eq_SINR}
{\rm{SINR}} =
\frac{{P_T}{G_{0,k}}h_{0,k}^{\left( {{S_0}} \right)}/{C_{0,k}^{\left( {{S_0}} \right)}}}{{\sigma_N^2} + \sum\nolimits_{S \in \left\{ {{\rm{LOS,NLOS}}} \right\}} {\sum\nolimits_{i \in \Phi _{{\rm{BS}}}^{\left( S \right)}\backslash {\rm{BS}}_0} \mathcal{I}_{i,k}^{\left( S, S_0 \right)} } }
\end{equation}
\noindent where $\mathcal{I}_{i,k}^{\left( S, S_0 \right)} = ({{P_T}{G_{i,k}}h_{i,k}^{\left( S \right)}/C_{i,k}^{\left( S \right)}){\bf 1}\left( {C_{i,k}^{\left( S \right)} > {C_{0,k}}^{\left( {{S_0}} \right)}} \right)}$ denotes the generic interfering term, ${\bf 1}\left(\cdot \right)$ denotes the indicator function that originates from the cell association, $S_0\in \left\{ {{\rm{LOS}},{\rm{NLOS}}} \right\}$ refers to the LOS/NLOS state of the BS$_0$-to-MT$_k$ link, ${\sigma_N^2}$ denotes the noise power, and the antenna gain of the BS$_0$-to-MT$_k$ link is $G_{0,k}=1$ for both omni-directional and 3GPP-based antenna radiation patterns. \vspace{-0.20cm}

In this paper, the performance metric for quantifying the accuracy of the PP-based model is the complementary cumulative distribution function of the SINR, since it provides complete information on its distribution. In addition, it corresponds to the coverage probability of a typical MT as a function of the link reliability threshold. Let $T$ be this threshold, it can be formulated as follows:
\begin{equation}
\label{Eq_Pcov}
{\rm{P}}_{{\mathop{\rm cov}} }\left( T \right) = \Pr \left\{ {{\rm{SINR}} > T} \right\} \vspace{-0.45cm}
\end{equation}
\begin{figure*}[!t]
\setcounter{equation}{12}
\begin{equation} \label{Eq_intensity_3GPP}
\begin{split}
& \Lambda _{{\rm{PL}}}^{\left( {{\rm{LOS,3GPP}}} \right)}\left( {\left[ {0,x} \right)} \right) = \mathcal{H}\left( {x - {\kappa _{{\rm{LOS}}}}r_0^{{\alpha _{{\rm{LOS}}}}}} \right)\left[ {{(1/2)}{{\left( {x/{\kappa _{{\rm{LOS}}}}} \right)}^{2/{\alpha _{{\rm{LOS}}}}}} \mathcal{\overline H} \left( {x - {\kappa _{{\rm{LOS}}}}{{18}^{{\alpha _{{\rm{LOS}}}}}}} \right)} \right.\\
&\left. { + \left( {624.064 - 36 {\exp \left( { - {{\left( {x/{\kappa _{{\rm{LOS}}}}} \right)}^{1/{\alpha _{{\rm{LOS}}}}}}/36} \right)}\left( {18 + {\left( {x/{\kappa _{{\rm{LOS}}}}} \right)^{1/{\alpha _{{\rm{LOS}}}}}}} \right) + 18 {\left( {x/{\kappa _{{\rm{LOS}}}}} \right)^{1/{\alpha _{\rm{LOS}}}}}} \right){\cal H}\left( {x - {\kappa _{{\rm{LOS}}}}{{18}^{{\alpha _{{\rm{LOS}}}}}}} \right)} \right] \\
&\Lambda _{{\rm{PL}}}^{\left( {\rm {NLOS,3GPP}} \right)}\left( {\left[ {0,x} \right)} \right) = {\cal H}\left( {x - {\kappa _{{\rm{NLOS}}}}{{18}^{{\alpha _{{\rm{NLOS}}}}}}} \right)\left[ {{(1/2)}{{\left( {{{\left( {x/{\kappa _{{\rm{NLOS}}}}} \right)}^{1/{\alpha _{{\rm{NLOS}}}}}} - 18} \right)}^2} - 786.064} \right.\\
& \hspace{2.87cm} \left. { + 36\exp \left( { - {(1/{36})}{{\left( {x/{\kappa _{{\rm{NLOS}}}}} \right)}^{1/{\alpha _{{\rm{NLOS}}}}}}} \right)\left( {18 + {{\left( {x/{\kappa _{{\rm{NLOS}}}}} \right)}^{1/{\alpha _{{\rm{NLOS}}}}}}} \right)} \right]
\end{split}
\end{equation}
\normalsize \hrulefill \vspace*{-15pt}
\end{figure*}
\begin{figure*}
\setcounter{equation}{13}
\begin{equation} \label{Eq_intensity_MultiBall}
\begin{split}
 \Lambda_{{\rm{PL}}} ^{\left( S, \rm{MultiBall} \right)} \left( {\left[ {0,x} \right)} \right) &= (1/2) q_{\left[ {0,d_1 } \right]}^{\left( S \right)} \left( {{x \mathord{\left/
 {\vphantom {x {\kappa ^{\left( S \right)} }}} \right.
 \kern-\nulldelimiterspace} {\kappa ^{\left( S \right)} }}} \right)^{{2 \mathord{\left/
 {\vphantom {2 {\alpha ^{\left( S \right)} }}} \right.
 \kern-\nulldelimiterspace} {\alpha ^{\left( S \right)} }}} {\mathcal{H}}\left( {x - \kappa ^{\left( S \right)} r_0^{\alpha ^{\left( S \right)} } } \right){\mathcal{\overline H}}\left( {x - \kappa ^{\left( S \right)} d_1^{\alpha ^{\left( S \right)} } } \right) \\
 & + (1/2) q_{\left[ {0,d_1 } \right]}^{\left( S \right)} d_1^2 {\mathcal{H}}\left( {x - \kappa ^{\left( S \right)} d_1^{\alpha ^{\left( S \right)} } } \right) + (1/2) q_{\left[ {d_N ,\infty } \right]}^{\left( S \right)} \left( {\left( {{x \mathord{\left/
 {\vphantom {x {\kappa ^{\left( S \right)} }}} \right.
 \kern-\nulldelimiterspace} {\kappa ^{\left( S \right)} }}} \right)^{{2 \mathord{\left/
 {\vphantom {2 {\alpha ^{\left( S \right)} }}} \right.
 \kern-\nulldelimiterspace} {\alpha ^{\left( S \right)} }}}  - d_N^2 } \right){\mathcal{H}}\left( {x - \kappa ^{\left( S \right)} d_N^{\alpha ^{\left( S \right)} } } \right) \\
 & + (1/2) \sum\nolimits_{n = 2}^N {q_{\left[ {d_{n - 1} ,d_n } \right]}^{\left( S \right)} \left( {\left( {{x \mathord{\left/
 {\vphantom {x {\kappa ^{\left( S \right)} }}} \right.
 \kern-\nulldelimiterspace} {\kappa ^{\left( S \right)} }}} \right)^{{2 \mathord{\left/
 {\vphantom {2 {\alpha ^{\left( S \right)} }}} \right.
 \kern-\nulldelimiterspace} {\alpha ^{\left( S \right)} }}}  - d_{n - 1}^2 } \right){\mathcal{H}}\left( {x - \kappa ^{\left( S \right)} d_{n - 1}^{\alpha ^{\left( S \right)} } } \right){\mathcal{\overline H}}\left( {x - \kappa ^{\left( S \right)} d_n^{\alpha ^{\left( S \right)} } } \right)}  \\
 & + (1/2) \sum\nolimits_{n = 2}^N {q_{\left[ {d_{n - 1} ,d_n } \right]}^{\left( S \right)} \left( {d_n^2  - d_{n - 1}^2 } \right){\mathcal{H}}\left( {x - \kappa ^{\left( S \right)} d_n^{\alpha ^{\left( S \right)} } } \right)}
\end{split}
\end{equation}
\normalsize \hrulefill \vspace*{-10pt}
\end{figure*}
\section{Tractable Simulation and Mathematical Modeling} \label{Approximations}
As mentioned in Section \ref{Introduction}, the PP-based model of cellular networks has been shown to be tractable under simplifying assumptions for the blockages, the path-loss functions and the antenna radiation patterns \cite{LopezPerez__LosNlos}, \cite{MDR_TWC_mmWave}, \cite{Singh__Globe2014}. In this section, we introduce some approximations for incorporating realistic models for the blockages and for the antenna radiation patterns in a PP-based modeling framework. To introduce and justify the proposed methodology, in particular, we assume that the BSs are distributed according to a PPP. The accuracy of the PP-based model is assessed in Section \ref{sec_simu_real}. The main focus of this section is, in fact, on proposing tractable approximations for modeling LOS and NLOS links that originate from realistic deployments of buildings, and for modeling practical antenna radiation patterns under the assumption that the BSs constitute a PPP.
\subsection{Tractable Modeling of Blockages} \label{Sec_ModelingBlockages}
Due to the presence of blockages (i.e., buildings), a generic BS-to-MT link may be either in LOS or in NLOS with a probability that depends on the BS-to-MT distance, the locations of the BSs, as well as the locations and the shapes of the buildings. This dependence on the distance and on the network topology makes the simulation and the analysis of cellular networks based on the PP-based model less tractable and more time-consuming. In this section, we introduce a piece-wise approximation for taking LOS and NLOS probabilities into account. The proposed approximation is optimized from the point of view of the typical MT and, thus, the spatial deployments of the BSs and of the buildings are explicitly taken into account. This makes it suitable for system-level performance evaluation and optimization. In this paper, it is referred to as the ``multi-ball approximation'' of blockages.
\vspace{-0.20cm}
\subsubsection{The Multi-Ball Model} \label{Sec_ModelingBlockages_MultiBall}
The idea behind the multi-ball approximation lies in replacing the actual LOS/NLOS probability of a typical MT with an approximated function, which still depends on the BS-to-MT distance $r$ but is piece-wise constant as a function of $r$. In particular, we split the BS-to-MT distance in $N+1$ regions, which correspond to $N$ balls whose center is at the MT. Let the radii of the $N$ balls be $d_1<d_2<,\ldots,<d_N$. Since $r$ is generic, i.e., $r \in \left[ {0, + \infty } \right)$, we have $d_1 \ge 0$ and $d_N <+ \infty$. The multi-ball approximation for the LOS/NLOS probability can be formulated as follows:
\setcounter{equation}{7}
\begin{equation}
\label{P_LOS_N_Ball}
p^{(S)}{\left( r\right)}=\sum\limits_{n = 1}^{N + 1} {q_{\left[ {{d_{n - 1}},{d_n}} \right]}^{(S)}{{\bf 1}_{\left[ {{d_{n - 1}},{d_n}} \right)}}} \left( r \right)
\end{equation}
\noindent where $S\in\left\{ {{\rm{LOS,NLOS}}} \right\}$, $d_0=0$, ${d_{N+1} }=+\infty$, ${{\bf{1}}_{\left[ {a,b} \right)}}\left( \cdot \right)$ is the generalized indicator function defined as ${{\bf{1}}_{\left[ {a,b} \right)}}\left( r \right)=1$ if $r\in\left[ {a,b} \right)$ and ${{\bf{1}}_{\left[ {a,b} \right)}}\left( r \right)=0$ if $r \notin \left[ {a,b} \right)$, and $q_{\left[ a,b \right]}^{(S)} \ge 0$ denotes the probability that a link of length $r\in\left[ {a,b} \right)$ is in state $S$. Since a link can be either in LOS or in NLOS, the following constraint need to be satisfied:
\begin{equation}
\sum\limits_{S \in \left\{ {{\rm{LOS}},{\rm{NLOS}}} \right\}}  {q_{\left[ {{d_0},{d_1}} \right]}^{(S)}}  =  \cdots  = \sum\limits_{S \in \left\{ {{\rm{LOS}},{\rm{NLOS}}} \right\}} {q_{\left[ {{d_N},{d_{N + 1}}} \right]}^{(S)}}  = 1 \vspace{-0.20cm}
\end{equation}

From \eqref{P_LOS_N_Ball}, it is apparent that the larger the number of balls $N$ is, the better the multi-ball approximation is. Simulation and mathematical complexity, however, increase as $N$ increases. In the next sections, we introduce a methodology for estimating $N$, $d_1<d_2<,\ldots,<d_N$, and $q_{\left[ {{d_{n - 1}},{d_n}} \right]}^{(S)}$ for $n=1, 2, \ldots, N$ and $S \in \left\{ {{\rm{LOS}},{\rm{NLOS}}} \right\}$ in order to find a good trade-off between complexity and accuracy.
\vspace{-0.20cm}
\subsubsection{Path-Loss Intensity Matching} \label{Sec_Matching}
The proposed approach for computing all the parameters in \eqref{P_LOS_N_Ball} is based on matching the intensities of the path-losses of the actual blockage model of interest and of the approximating blockage model in \eqref{P_LOS_N_Ball}. More specifically, the proposed approach leverages the displacement theorem of PPPs \cite{Blaszczyszyn_Infocom2013}. \vspace{-0.45cm}

Let $r_{i,k}$ be the link distance from a generic BS, BS$_i$, to the probe MT, MT$_k$. From Section \ref{Channel}, the path-loss is $l^{\left( S \right)} \left( r_{i,k} \right) = \kappa ^{\left( S \right)} \left( {\max \left\{ {r_0 ,r_{i,k}} \right\}} \right)^{\alpha^{\left( S \right)}}$ for $S \in \left\{ {{\rm{LOS}},{\rm{NLOS}}} \right\}$. By assuming that the BSs are distributed according to a PPP of density $\lambda _{{\rm{BS}}}$, it can be proved that the PP of the path-losses $\left\{ {\kappa ^{\left( S \right)} \left( {\max \left\{ {r_0 ,r_{i,k} } \right\}} \right)^{\alpha ^{\left( S \right)} } ,i \in \Phi _{{\rm{BS}}}^{\left( S \right)} } \right\}$ is still a PPP, whose intensity measure can be formulated as follows:
\begin{equation} \label{Intensity_1}
\Lambda _{{\rm{PL}}} \left( {\left[ {0,x} \right)} \right) = 2\pi \lambda _{{\rm{BS}}} \sum\nolimits_{S \in \left\{ {{\rm{LOS}},{\rm{NLOS}}} \right\}} {\Lambda _{{\rm{PL}}}^{\left( S \right)} \left( {\left[ {0,x} \right)} \right)}
\end{equation}
\begin{equation} \label{Intensity_2}
\begin{split}
 & \Lambda _{{\rm{PL}}}^{\left( S \right)} \left( {\left[ {0,x} \right)} \right) = \int\nolimits_0^{ + \infty } {\Pr \left\{ {l^{\left( S \right)} \left( r \right) \le x} \right\}p^{\left( S \right)} \left( r \right)rdr}  \\
  & = \int\nolimits_0^{ + \infty } {\Pr \left\{ {\kappa ^{\left( S \right)} \left( {\max \left\{ {r_0 ,r} \right\}} \right)^{\alpha ^{\left( S \right)} }  \le x} \right\}p^{\left( S \right)} \left( r \right)rdr}
 \end{split} \vspace{-0.20cm}
\end{equation}

\vspace{-0.20cm}
Equation \eqref{Intensity_1} is a direct application of the displacement theorem of PPPs. Further details are available in \cite{Blaszczyszyn_Infocom2013}. \vspace{-0.20cm}

The intensity measure in \eqref{Intensity_1} is uniquely determined by the probabilities ${p^{\left( \rm{LOS} \right)} \left( \cdot \right)}$ and ${p^{\left( \rm{NLOS} \right)} \left( \cdot \right)}$ as a function of the link distance $r$. Let $\Lambda _{{\rm{PL}}}^{\left( {{\rm{actual}}} \right)} \left( {\left[ { \cdot , \cdot } \right)} \right)$ be the intensity measure that corresponds to the actual probabilities of LOS and NLOS. For example, it can be obtained by using the empirical- and the 3GPP-based blockage models in Sections \ref{LOS_NLOS__Empirical} and \ref{LOS_NLOS__3GPP}. Let $\Lambda _{{\rm{PL}}}^{\left( {{\rm{approx}}} \right)} \left( {\left[ { \cdot , \cdot } \right)} \right)$ be the intensity measure that corresponds to the multi-ball approximation in \eqref{Intensity_1}. We propose to estimate $N$, $d_1<d_2<,\ldots,<d_N$, and $q_{\left[ {{d_{n - 1}},{d_n}} \right]}^{(S)}$ for $n=1, 2, \ldots, N$ and $S \in \left\{ {{\rm{LOS}},{\rm{NLOS}}} \right\}$ in \eqref{Intensity_1} by minimizing the following utility error function:
\begin{equation} \label{Eq_matching}
\left\| {\ln \left( {\Lambda _{{\rm{PL}}}^{\left( {{\rm{actual}}} \right)} \left( {\left[ {0,x_{M } } \right)} \right)} \right) - \ln \left( {\Lambda _{{\rm{PL}}}^{\left( {{\rm{approx}}} \right)} \left( {\left[ {0,x_{M } } \right)} \right)} \right)} \right\|_F^2
\end{equation}
\noindent where ${\left\|  \cdot  \right\|_F}$ denotes the Frobenius norm and $x_M$ is chosen in order to capture the main body of $\Lambda _{{\rm{PL}}}^{\left( {{\rm{actual}}} \right)} \left( {\left[ { \cdot , \cdot } \right)} \right)$, i.e., it is close to zero for $x>x_M$. The rationale of \eqref{Eq_matching} originates from the fact that, from the point of view of a typical MT, the impact of blockages is almost the same if the intensity measures are close to each other. The logarithm in \eqref{Eq_matching} is used to make the approximation more accurate. \vspace{-0.20cm}

In practice, the optimization problem in \eqref{Eq_matching} can be solved by using, e.g., the function $\mathsf{lsqcurvefit}$, which is built-in in Matlab. We have verified that the solution is quite stable with respect to the choice of the initial point of the search, which, then, can be chosen at random. The computation of \eqref{Eq_matching}, however, requires the closed-form expressions of the intensity measures $\Lambda _{{\rm{PL}}}^{\left( {{\rm{actual}}} \right)} \left( {\left[ { \cdot , \cdot } \right)} \right)$ and $\Lambda _{{\rm{PL}}}^{\left( {{\rm{approx}}} \right)} \left( {\left[ { \cdot , \cdot } \right)} \right)$. They are provided in what follows for relevant case studies. \vspace{-0.55cm}
\paragraph{Empirical-Based Model} The intensity measure is computed based on the actual footprints of the buildings, according to Section \ref{LOS_NLOS__Empirical}. Since LOS and NLOS probabilities are not available in closed-form in this case, \eqref{Intensity_2} cannot be directly applied and the intensity measure cannot be formulated in closed-form. For ease of description, the procedure for computing it is discussed in Section \ref{sec_simu_real}. \vspace{-0.55cm}
\paragraph{3GPP-Based Model} The intensity measure is computed by assuming the LOS and NLOS probabilities in Section \ref{LOS_NLOS__3GPP}. More specifically, by inserting \eqref{3GPP_LOS} in \eqref{Intensity_2} and computing the integrals, the intensity measure in \eqref{Eq_intensity_3GPP} is obtained, where ${\cal H}\left(  \cdot  \right)$ is the Heaviside function and $\overline {\cal H} \left(  x  \right) = 1 - {\cal H}\left(  x  \right)$. The result in \eqref{Eq_intensity_3GPP} holds under the assumption $r_0 < 18$ meters, which is usually satisfied for typical setups. \vspace{-0.55cm}
\paragraph{Multi-Ball Model} Similar to the 3GPP-based model, the intensity measure can be formulated in closed-form by inserting \eqref{P_LOS_N_Ball} in \eqref{Intensity_2} and by computing the resulting integrals. The final result is available in \eqref{Eq_intensity_MultiBall} for $r_0 < d_1$, which usually holds for typical setups. \vspace{-0.55cm}
\paragraph{1-State Model} The 1-state model can be viewed as a special case of the multi-ball model, where $N=0$, $d_0=0$, $d_1=\infty$, as well as $q_{\left[ {0,\infty } \right]}^{\left( \rm{LOS} \right)}  = 1$ and $q_{\left[ {0,\infty } \right]}^{\left( \rm{NLOS} \right)}  = 0$ or $q_{\left[ {0,\infty } \right]}^{\left( \rm{LOS} \right)}  = 0$ and $q_{\left[ {0,\infty } \right]}^{\left( \rm{NLOS} \right)}  = 1$. Thus, the intensity measure follows from \eqref{Eq_intensity_MultiBall}, and it can be formulated as ($S=\rm{LOS}$ or $S=\rm{NLOS}$):
\setcounter{equation}{14}
\begin{equation} \label{Eq_intensity_1state}
\Lambda _{{\rm{PL}}}^{({\rm{1state}})} \left( {\left[ {0,x} \right)} \right) = \pi \lambda _{{\rm{BS}}} \left( {{x \mathord{\left/
 {\vphantom {x {\kappa ^{\left( S \right)} }}} \right.
 \kern-\nulldelimiterspace} {\kappa ^{\left( S \right)} }}} \right)^{{2 \mathord{\left/
 {\vphantom {2 {\alpha ^{\left( S \right)} }}} \right.
 \kern-\nulldelimiterspace} {\alpha ^{\left( S \right)} }}} {\mathcal{H}}\left( {x - \kappa ^{\left( S \right)} r_0^{\alpha ^{\left( S \right)} } } \right) \vspace{-0.35cm}
\end{equation}
\begin{table*}[!t]
\centering
\caption{3-ball approximation of empirical and 3GPP blockage models obtained as the solution of \eqref{Eq_matching}.}
\begin{tabular}{|c|c|c|c|c|c|c|c|} \hline
 &$d_1$ (meters) & $d_2$ (meters) & $d_3$ (meters) & $q_{\left[ {{0},{d_1}} \right]}^{(\rm LOS)}$ & $q_{\left[ {{d_1},{d_2}} \right]}^{(\rm LOS)}$ & $q_{\left[ {{d_2},{d_3}} \right]}^{(\rm LOS)}$ & $q_{\left[ {{d_3},{\infty }} \right]}^{(\rm LOS)}$\\ \hline
London&15.1335&56.5978&195.7149&0.7948&0.3818&0.0939&0\\ \hline
Manchester&13.2076&57.8840&213.3940&0.7866&0.4981&0.1015&0.0001\\ \hline
3GPP&47.7989&215.9387&1874.442&0.9446&0.2142&0.0243&0.0021\\
\hline\end{tabular}
\label{Table3} \vspace{-0.10cm}
\end{table*}
\subsection{Tractable Modeling of Radiation Patterns}
The antenna radiation pattern plays an important role for system-level performance evaluation. Incorporating practical antenna radiation patterns in system-level simulations and mathematical analysis may, however, not be straightforward, as it usually entails a loss of tractability. This is the reason why the omni-directional radiation pattern in Section \ref{Sec_OmniAntenna} is usually adopted for system-level analysis. In this section, we introduce a tractable approximation for modeling arbitrary antenna radiation patterns, which can be readily used for system-level analysis and simulation. \vspace{-0.20cm}

Let $g^{\left( {{\rm{actual}}} \right)} \left(  \cdot  \right)$ be the actual antenna radiation pattern of interest. We propose to approximate it by using a multi-lobe antenna radiation pattern, where the antenna gain is constant in each lobe. In this paper, it is referred to as ``multi-lobe approximation''. In mathematical terms, the approximating antenna radiation pattern can be formulated as $g^{\left( {{\rm{approx}}} \right)} \left( \theta  \right) = g^{\left( {{\rm{MultiLobe}}} \right)} \left( \theta  \right)$, where:
\begin{equation}
\label{Eq_multilobe}
g^{\left( {{\rm{MultiLobe}}} \right)} \left( \theta  \right) = { \begin{cases}
{g^{(1)}} & \quad \textrm {if}\; {\left| \theta  \right| \le \theta^{\left( 1 \right)}}\\
{g^{(2)}} & \quad \textrm {if}\; {\theta^{\left( 1 \right)}<\left| \theta  \right| \le \theta^{\left( 2 \right)}} \\
 \vdots  & \quad  \vdots\\
{g^{(K)}} & \quad \textrm {if}\; {\theta^{\left( K-1 \right)}  < \left| \theta  \right| \le \pi }\\
 \end{cases} }
\end{equation}
\noindent with $K$ denoting the number of lobes and $0< \theta^{\left( 1 \right)}<\ldots< \theta^{\left( K-1 \right)}<\pi$ being the angles that correspond to the $K$ lobes. \vspace{-0.40cm}

Similar to \eqref{Eq_matching}, the parameters of the multi-lobe approximation in \eqref{Eq_multilobe}, i.e., $K$ and $0< \theta^{\left( 1 \right)}<\ldots< \theta^{\left( K-1 \right)}<\pi$, are computed by minimizing the following utility error function:
\begin{equation}
\label{Eq_matching_pattern}
{\left\| {{{\log }_{10}}\left( {g^{\left( {{\rm{actual}}} \right)}\left( \theta  \right)} \right) - {{\log }_{10}}\left( {g^{\left( {{\rm{approx}}} \right)}\left( \theta  \right)} \right)} \right\|_F^2} \vspace{-0.20cm}
\end{equation}

Similar to \eqref{Eq_matching}, the logarithm is used for guaranteeing a better approximation. Since the antenna radiation pattern is usually measured in dB, ${{\log }_{10}}(\cdot)$ instead of ${{\ln}}(\cdot)$ is used. It is apparent from \eqref{Eq_multilobe} that the larger the number of lobes $K$ is, the better the approximation is. The numerical complexity increases, however, as $K$ increases. In Section \ref{sec_simu_real}, numerical examples confirm that a good accuracy can be obtained even by using a small number of lobes.
\section{Numerical Results: Experimental Validation}\label{sec_simu_real}
In this section, we illustrate several numerical examples in order to validate the accuracy of the PP-based abstraction for modeling cellular networks, as well as to confirm the tightness of the proposed multi-ball and multi-lobe approximations for simplifying the simulation and for enabling the mathematical modeling of cellular networks.
\vspace{-0.45cm}
\paragraph{Simulation Setup} Unless otherwise stated, the following simulation setup, which is in agreement with the long term evolution advanced (LTE-A) standard, is assumed. The transmit power of the BSs is $P_T=30$ dBm; the noise power is $\sigma_N^2=-174+10\log_{10}\left( {{\rm{B}_{\rm W}}} \right)+{\cal F}_{\rm dB}$, where ${\rm B_W}$=20 MHz is the transmission bandwidth and ${\cal F}_{\rm dB}=10$ dB is the noise figure; $\kappa_{\rm LOS}=\kappa_{\rm NLOS}={\left( {4\pi /\nu } \right)^2}$ is the free space path-loss at a distance of 1 meter from the transmitter, where $\upsilon  = c/{f_c}$ is the signal wavelength, $c \approx 3 \times {10^8}$ meters/sec is the speed of light, and $f_c$=2.1 GHz is the signal frequency; $r_0=1$ meter; the path-loss exponents of LOS and NLOS links are $\alpha_{\rm LOS}=2.5$ and $\alpha_{\rm NLOS}=3.5$; the mean and standard deviation of the shadowing are ${\mu _{{\rm{LOS}}}} = {\mu _{{\rm{NLOS}}}} = 0$ dB, ${\sigma _{{\rm{LOS}}}} = 5.8$ dB, and ${\sigma _{{\rm{NLOS}}}} = 8.7$ dB; the fast-fading envelope of the LOS links follows a Nakagami-$m$ distribution with parameters $m=2$ and ${\Omega _{\rm LOS}} = 1$; the fast-fading envelope of the NLOS links follows a Rayleigh distribution with parameter ${\Omega _{\rm NLOS}} = 1$. Used notation: ``O2'' means that only the BSs from O2 are accessible; ``Vodafone'' means that only the BSs from Vodafone are accessible; and ``O2+Vodafone'' means that all BSs from O2 and Vodafone are accessible.
\begin{figure}[!t]
\flushleft
\includegraphics [width=\columnwidth] {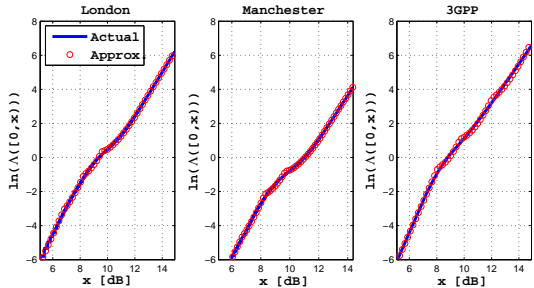}
\vspace{-0.75cm} \caption{Comparison of the intensity measures of empirical- and 3GPP-based models against their multi-ball ($N=3$) approximation counterpart.}
\label{Fig_intensity} \vspace{-0.15cm}
\end{figure}
\vspace{-0.45cm}
\paragraph{Multi-Ball Approximation of the Blockages} In Fig. \ref{Fig_intensity}, we study the accuracy of the proposed multi-ball approximation for modeling spatial blockages, under the assumption that the BSs are distributed according to a PPP. More specifically, Fig. \ref{Fig_intensity} compares the intensity measure of the empirical-based and of the 3GPP-based models against their multi-ball approximations, which are obtained by solving the optimization in \eqref{Eq_matching}. The solution of the optimization is summarized in Table \ref{Table3} by assuming $N=3$, which provides a good matching accuracy while still keeping the computational complexity at a low level. The intensity measure of the empirical-based model of the blockages is obtained by using the following procedure based on the actual locations and shapes of the buildings obtained from the OS database. \vspace{-0.55cm}
\begin{description}
  \item[Step 1:] The geographical regions illustrated in Figs. \ref{Fig_London} and \ref{Fig_Manchester} for London and Manchester are considered. As discussed in Section \ref{sec_Building_distribution}, this data is obtained from OS. The BSs are generated according to a PPP of density $\lambda_{\rm BS}$, which is chosen according to the data in Tables \ref{Table1} and \ref{Table2}. Outdoor and rooftop BSs are identified. \vspace{-0.25cm}
  \item[Step 2:] In the same areas, the MTs are generated according to another PPP of density $\lambda_{\rm MT}=10\lambda_{\rm BS}$. This choice of $\lambda_{\rm MT}$ guarantees saturated traffic conditions, i.e., all the BSs have at least one MT to serve based on the cell association in Section \ref{Sec_association}. Among all the MTs, one MT among those that do not lay in a building (outdoor MTs) is randomly chosen as the typical MT. \vspace{-0.55cm}
  \item[Step 3:] Let the probe (typical) MT, its distance $r$ and link state (LOS or NLOS) with respect to any accessible BSs are computed according to  Section \ref{LOS_NLOS__Empirical}. \vspace{-0.25cm}
  \item[Step 4:] Step 2 and Step 3 are repeated several thousands of times in order to get sufficient statistical data. From this data, two vectors are obtained: a vector containing the distances whose links are in LOS and a vector containing the distances whose links are in NLOS. \vspace{-0.25cm}
  \item[Step 5:] From the vectors computed in Step 4, $p^{\left( {{\rm{LOS}}} \right)} \left( r \right)$ and $p^{\left( {{\rm{NLOS}}} \right)} \left( r \right)$ are estimated by using, e.g., the $\mathsf{hist}$ function of Matlab. To this end, a resolution step equal to $\Delta r=1$ meter and $M_t=2000$ discrete distances are considered. Thus, the LOS and NLOS probabilities are available for the set of distances $r_t$ for $t=1,2,\ldots,M_t$, where $\Delta r=r_t - r_{t-1} = 1$ meter. The corresponding LOS and NLOS probabilities are $p^{\left( {{\rm{LOS}}} \right)} \left( r_t \right)$ and $p^{\left( {{\rm{NLOS}}} \right)} \left( r_t \right)$ for $t=1,2,\ldots,M_t$. \vspace{-0.25cm}
  \item[Step 6:] Finally the intensity measure of the path-losses is computed by using \eqref{Intensity_1} and the following discrete (empirical) approximation of \eqref{Intensity_2} (for $S = \left\{ {{\rm{LOS}},{\rm{NLOS}}} \right\}$):
\begin{equation} \label{Intensity_Empirical}
\begin{split}
& \hspace{-0.80cm} \Lambda _{{\rm{PL}}}^{\left( S \right)} \left( {\left[ {0,x} \right)} \right) \\
& \hspace{-0.80cm} \approx \Delta r\sum\limits_{t = 1}^{M_t } {\Pr \left\{ {\kappa ^{\left( S \right)} \left( {\max \left\{ {r_0 ,r_t } \right\}} \right)^{\alpha ^{\left( S \right)} }  \le x} \right\}p^{\left( S \right)} \left( {r_t } \right)r_t }
\end{split}
\end{equation}
\end{description}

\vspace{-0.25cm}
The empirical intensity measure in \eqref{Intensity_Empirical} constitute the ``actual'' intensity for the left and middle plots in Fig. \ref{Fig_intensity}. The right plot is obtained by assuming \eqref{Eq_intensity_3GPP} as the ``actual'' intensity. All in all, a good accuracy is obtained.
\begin{figure}[!t]
\centering
\includegraphics [width=0.90\columnwidth] {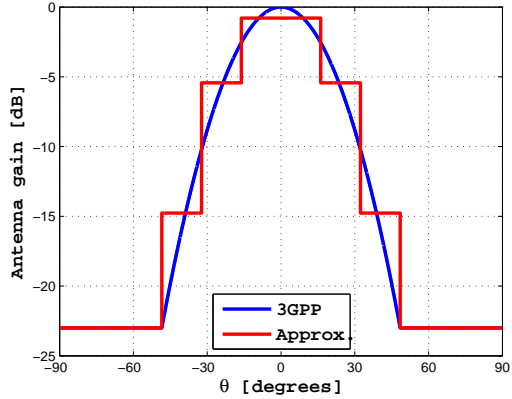}
\vspace{-0.35cm} \caption{3GPP-based antenna radiation pattern in \eqref{G_3GPP} and its multi-lobe approximation obtained by using \eqref{Eq_multilobe} and \eqref{Eq_matching_pattern}. The multi-lobe approximation is obtained for $K=4$ and the solution of \eqref{Eq_matching_pattern} is $g^{(1)}=0.8341$,  $g^{(2)}=0.2865$, $g^{(3)}=0.0334$, $g^{(4)}=0.005$, $\theta^{(1)}=16.152^\circ$, $\theta^{(2)}=32.304^\circ$, and $\theta^{(3)}=48.455^\circ$.}
\label{Fig_pattern} \vspace{-0.35cm}
\end{figure}
\vspace{-0.45cm}
\paragraph{Multi-Lobe Approximation of the Antenna Radiation Patterns} In Fig. \ref{Fig_pattern}, we study the accuracy of the proposed multi-lobe approximation for modeling realistic antenna radiation patterns. In particular, we test the multi-lobe approximation for modeling the 3GPP-based antenna radiation pattern in \eqref{G_3GPP}. The figure shows that a good accuracy is obtained even though $K=4$. The approximation error can be reduced by increasing $K$. In what follows, we show, however, that the 4-lobe approximation in Fig. \ref{Fig_pattern} is sufficiently accurate for system-level performance analysis, while offering a low computational complexity.
\begin{figure}[!t]
\centering
\includegraphics [width=\columnwidth] {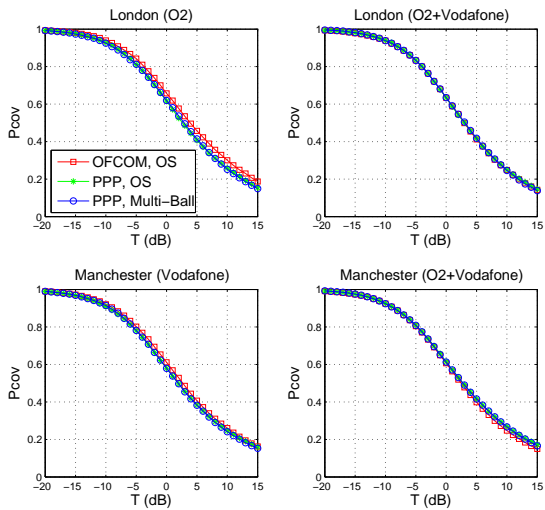}
\vspace{-1.00cm} \caption{Coverage probability by assuming an omni-directional antenna radiation pattern. Three case studies are analyzed. 1) ``OFCOM, OS'': the BSs are obtained from the OFCOM database and the buildings from the OS database. 2) ``PPP, OS'': the BSs are distributed according to a PPP and the buildings are obtained from the OS database. 3) ``PPP, Multi-Ball'': the BSs are distributed according to a PPP and the multi-ball approximation in Table \ref{Table3} (London and Manchester) is used.}
\label{Fig_real_omni} \vspace{-0.35cm}
\end{figure}
\begin{figure}[!t]
\centering
\includegraphics [width=\columnwidth] {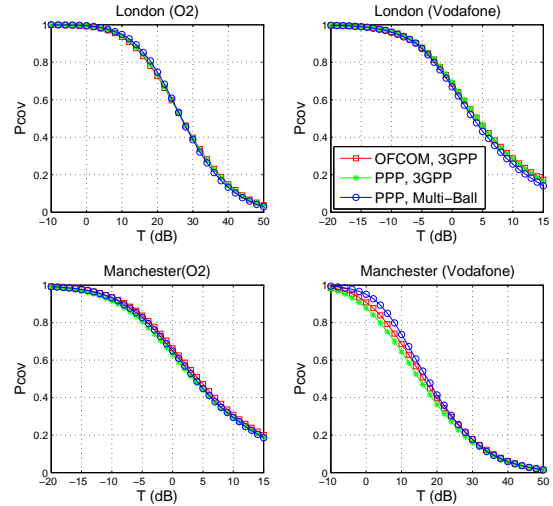}
\vspace{-1.00cm} \caption{Coverage probability by assuming an omni-directional antenna radiation pattern. Three case studies are analyzed. 1) ``OFCOM, 3GPP'': the BSs are obtained from the OFCOM database and the LOS/NLOS link states are obtained from \eqref{3GPP_LOS}. 2) ``PPP, 3GPP'': the BSs are distributed according to a PPP and the LOS/NLOS link states are obtained from \eqref{3GPP_LOS}. 3) ``PPP, Multi-Ball'': the BSs are distributed according to a PPP and the multi-ball approximation in Table \ref{Table3} (3GPP) is used.}
\label{Fig_3GPP_pathloss} \vspace{-0.35cm}
\end{figure}
\begin{figure}[!t]
\centering
\includegraphics [width=\columnwidth] {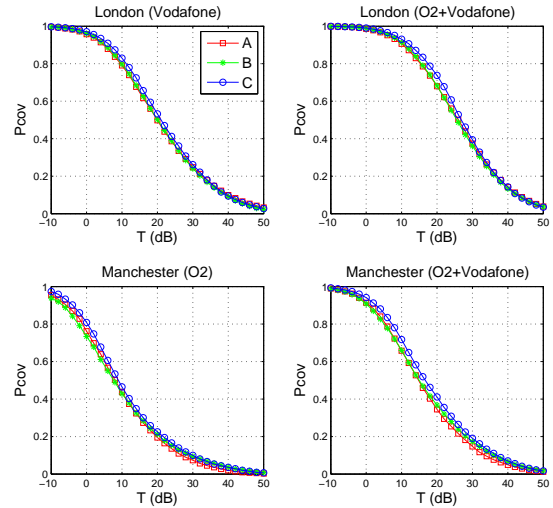}
\vspace{-1.00cm} \caption{Coverage probability by assuming the 3GPP antenna radiation pattern in \eqref{G_3GPP}. Three case studies are analyzed. 1) ``A'': the BSs are obtained from the OFCOM database and the buildings from the OS database. 2) ``B'': the BSs are distributed according to a PPP and the buildings are obtained from the OS database. 3) ``C'': the BSs are distributed according to a PPP, the multi-ball approximation in Table \ref{Table3} (London and Manchester) is used, and the 3GPP antenna radiation pattern is replaced by its multi-lobe approximation in Fig. \ref{Fig_pattern}.}
\label{Fig_real_3GPP} \vspace{-0.35cm}
\end{figure}
\vspace{-0.45cm}
\paragraph{PP-Based Modeling of the BSs} In Figs. \ref{Fig_real_omni}-\ref{Fig_real_3GPP}, we study the accuracy of the PP-based abstraction for modeling cellular networks, by either considering or not the multi-ball and the multi-lobe approximations for modeling blockages and antenna radiation patterns, respectively. As discussed in Sections \ref{sec_BS_distribution} and \ref{sec_Building_distribution}, the empirical coverage probability is obtained by using the locations of the BSs and the footprints of the buildings from the OFCOM and OS databases, respectively. The following procedure for computing the empirical coverage probability is used. \vspace{-0.55cm}
\begin{description}
  \item[Step 1:] The geographical regions illustrated in Figs. \ref{Fig_London} and \ref{Fig_Manchester} for London and Manchester are considered. As discussed in Section \ref{sec_Building_distribution}, this data is obtained from OS. Two case studies for the locations of the BSs are considered. 1) The BSs are distributed according to their actual locations obtained from OFCOM (Figs. \ref{Fig_London}, \ref{Fig_Manchester}, Tables \ref{Table1}, \ref{Table2}). 2) The BSs are distributed according to a PPP whose density is the same as that of Tables \ref{Table1}, \ref{Table2}. In both cases, outdoor and rooftop BSs are identified. \vspace{-0.55cm}
  \item[Step 2:] In the same areas, the MTs are generated according to another PPP of density $\lambda_{\rm MT}=10\lambda_{\rm BS}$. This choice of $\lambda_{\rm MT}$ guarantees saturated traffic conditions, i.e., all the BSs have at least one MT to serve based on the cell association in Section \ref{Sec_association}. Among all the MTs, one MT among those that do not lay in a building (outdoor MTs) is randomly chosen as the typical MT. \vspace{-0.55cm}
  \item[Step 3:] Let the probe (typical) MT, its distance $r$ and link state (LOS or NLOS) with respect to any accessible BSs are computed according to Section \ref{LOS_NLOS__Empirical}. \vspace{-0.25cm}
  \item[Step 4:] For each link between the probe MT and the accessible BSs, path-loss, shadowing and fast-fading gains are generated according to Section \ref{Channel}. \vspace{-0.25cm}
  \item[Step 5:] Let the probe MT and the accessible BSs, its serving BS is identified by using \eqref{Eq_C} in Section \ref{Sec_association}. \vspace{-0.25cm}
  \item[Step 6:] The antenna gain of the probe link is set equal to one, while the antenna gains of all the other links are generated as described in Sections \ref{Channel} and \ref{Sec_antenna}. \vspace{-0.25cm}
  \item[Step 7:] The SINR and its associated coverage probability are computed by using \eqref{Eq_SINR} and \eqref{Eq_Pcov}, respectively. \vspace{-0.25cm}
  \item[Step 8:] Steps 1-7 are repeated $10^6$ times in order to get sufficient statistical data. The final coverage probability is computed as the empirical mean of the obtained $10^6$ realizations for each target reliability threshold.
\end{description}

\vspace{-0.20cm}
If the LOS and NLOS probabilities are computed based on the 3GPP-based blockage model in Section \ref{LOS_NLOS__3GPP}, the same procedure is used. The only difference is that the actual locations of the buildings are ignored and each link state (LOS vs. NLOS) is identified, according to \eqref{3GPP_LOS}, only based on the distance between the probe MT and each accessible BS. The same comment applies to the 1-state model in Section \ref{LOS_NLOS__1state}. In this case, all the links are assumed to be, a priori, either in LOS or NLOS. \vspace{-0.20cm}

All in all, Figs. \ref{Fig_real_omni}-\ref{Fig_real_3GPP} confirm the accuracy of the PP-based abstraction model, and the tightness of the proposed multi-ball and multi-lobe approximations in practical scenarios.
\begin{figure}[!t]
\centering
\includegraphics [width=\columnwidth] {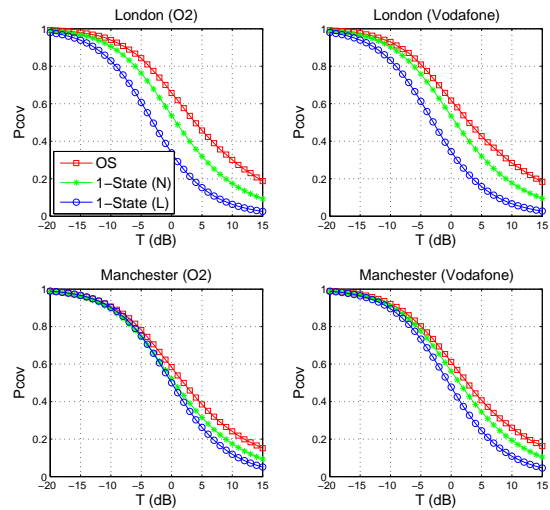}
\vspace{-1.00cm} \caption{Coverage probability: impact of blockages (omni-directional antennas). Three case studies are analyzed. 1) ``OS'': the buildings are obtained from the OS database. 2) ``1-State (N)'': all links are in NLOS. 3) ``1-State (L)'': all links are in LOS.}
\label{Fig_compare_state} \vspace{-0.35cm}
\end{figure}
\begin{figure}[!t]
\centering
\includegraphics [width=\columnwidth] {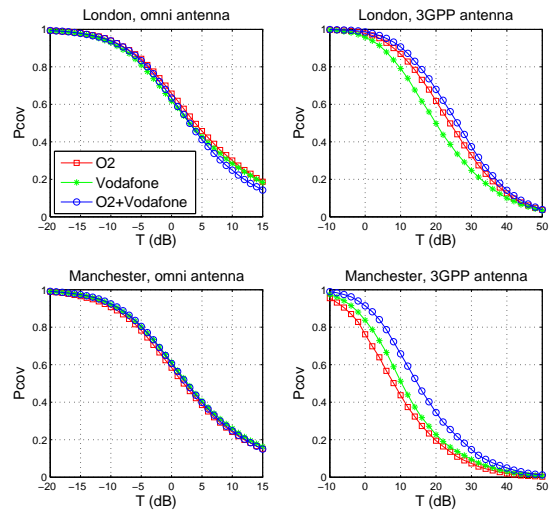}
\vspace{-1.00cm} \caption{Coverage probability: impact of the antenna radiation pattern (the buildings are obtained from the OS database, i.e., empirical-based blockages are considered). Three case studies are analyzed. 1) ``O2'': only the BSs from O2 are accessible. 2) ``Vodafone'': only the BSs from Vodafone are accessible. 3) ``O2+Vodafone'': all BSs from O2 and Vodafone are accessible.}
\label{Fig_O2_Vod_Both} \vspace{-0.35cm}
\end{figure}
\vspace{-0.55cm}
\paragraph{Achievable Performance: Impact of Blockages and Antenna Radiation Patterns} In Fig. \ref{Fig_compare_state}, we study the impact of the blockage model on the coverage probability. This figure highlights the importance of accurately modeling blockages. More specifically, the widespread used 1-state model provides different results from the more accurate and realistic LOS/NLOS blockage model, which accounts for the locations of buildings. Figure \ref{Fig_compare_state} points out that the coverage probability may be better than that predicted by using the 1-state model, since the links in LOS result in good probe links while the links in NLOS result in less interference. The proposed multi-ball approximation turns out to be a useful tool for taking LOS/NLOS propagation conditions into account at an affordable complexity, eventually leading to the mathematical analysis and optimization of cellular networks. \vspace{-0.20cm}

In Fig. \ref{Fig_O2_Vod_Both}, we study the impact of the antenna radiation pattern on the coverage probability. This figure highlights the importance of accurately modeling the antenna radiation pattern. In particular, it shows that sharing the infrastructure between telecommunication operators may not result in any improvements of the coverage probability if omni-directional antennas are used. If, on the other hand, practical (i.e., directional) antennas are used, e.g., based on 3GPPP recommendations (see \eqref{G_3GPP}), the coverage probability may be better due to the denser deployment of BSs. This is due to the fact that directional antennas have the inherent capability of reducing the impact of interference, which is desirable in interference-limited cellular networks.
\section{Conclusion} \label{Conclusion}
With the aid of experimental data for the actual locations of BSs and for the actual locations and shapes of buildings, we have studied the accuracy of the PP-based abstraction for modeling cellular networks. This study has highlighted that the PP-based model is sufficiently accurate for modeling dense urban environments of major metropolitan areas. We have observed that accurate models for the blockages and for the antenna radiation patterns are needed for obtaining reliable estimates of the coverage probability of cellular networks. Finally, we have proposed flexible approximations for incorporating realistic blockage and antenna models into the PP-based abstraction of cellular networks, and have validated their accuracy in relevant scenarios. Based on these findings, the PP-based model seems to be sufficiently accurate and tractable for enabling the mathematical analysis and optimization of emerging ultra-dense cellular networks, which use advanced wireless access transmission schemes.
\section{Acknowledgment}
This work was supported in part by the European Commission through the FP7-PEOPLE MITN-CROSSFIRE Project under Grant 317126 and through the H2020-PEOPLE ETN-5Gwireless Project under Grant 641985.

\bibliographystyle{unsrt}%if we want the reference ordered by the citing sequence, use {unsrt}
\end{document}